\documentclass[aps,prl,twocolumn,tworchcolumn, showpacs,preprintnumbers,amsmath,amssymb,superscriptaddress,floatfix]{revtex4}

\usepackage{graphicx}
\usepackage[usenames]{color}

\newcommand{\be}{\begin{equation}}
\newcommand{\ee}{\end{equation}}
\newcommand{\bi}{\begin{itemize}}
\newcommand{\ei}{\end{itemize}}
\newcommand{\ben}{\begin{enumerate}}
\newcommand{\een}{\end{enumerate}}

\newcommand{\unit}{\hat{\bf n}}

\newcommand{\rv}{{\bf r}}

\newcommand{\dv}{{\bf d}}

\newcommand{\qv}{{\bf q}}
\newcommand{\kv}{{\bf k}}
\newcommand{\eo}{\epsilon_0}

\newcommand{\beq}{\begin{equation}}
\newcommand{\eeq}{\end{equation}}
\newcommand{\bea}{\begin{eqnarray}}
\newcommand{\eea}{\end{eqnarray}}

\renewcommand{\(}{\left(}
\renewcommand{\)}{\right)}

\newcommand{\commentout}[1]{{}}

\newcommand{\eq}[1]{Eq.~\eqref{#1}}

\begin{document}

\title{Light scattering for thermometry of fermionic atoms in an optical lattice}

\author{J. Ruostekoski}
\affiliation{School of Mathematics, University of Southampton,
Southampton, SO17 1BJ, United Kingdom}
\author{C. J. Foot}
\affiliation{Clarendon Laboratory, University of Oxford, Parks Road,
Oxford, OX1 3PU, UK}
\author{A. B. Deb}
\affiliation{Clarendon Laboratory, University of Oxford, Parks Road,
Oxford, OX1 3PU, UK}
\date{\today}

\begin{abstract}
We propose a method for measuring the temperature of fermionic atoms
in an optical lattice potential from the intensity of the
scattered light in the far-field diffraction pattern. We consider a
single-component gas in a tightly-confined two-dimensional lattice,
illuminated by far off-resonant light driving a cycling transition.
Our calculations show that thermal correlations of the fermionic atoms generate fluctuations in 
the intensity of the diffraction pattern of light scattered from the atomic lattice array and 
that this signal can be accurately detected above the shot noise using a lens to collect 
photons scattered in a forward direction (with the diffraction maxima blocked).
The sensitivity of the thermometer is enhanced by an additional
harmonic trapping potential.
\end{abstract}


\maketitle

Ultra-cold atomic gases in optical lattices can constitute almost ideal realizations
of Hubbard models \cite{JAK98,GRE02a} that are fundamental to strongly-correlated physics.
Recent experiments on fermionic atoms in lattices have demonstrated both a superfluid pairing 
\cite{CHI06} and a Mott insulator \cite{JOR08,SCH08}, opening up possibilities
for experimental simulation of even more complex strongly-correlated systems, such as
antiferromagnetic phases and high-$T_c$ superconductivity.
Thermal energy is a major
control parameter in fermionic lattice systems and characterizing  phase diagrams
is fundamentally related to the ability to perform accurate
temperature measurements.

Here we show that the temperature of fermionic atoms in a 2D optical lattice can
be accurately determined by measurement of the light scattering from the atoms.
The diffraction pattern is insensitive to thermal atomic correlations but blocking the diffraction
maxima and collecting light scattered outside the diffraction orders using a lens 
provides a measurable optical signal that reflects
thermal and quantum fluctuations of lattice atoms.
\begin{figure}[t]
\includegraphics[width=0.8\columnwidth]{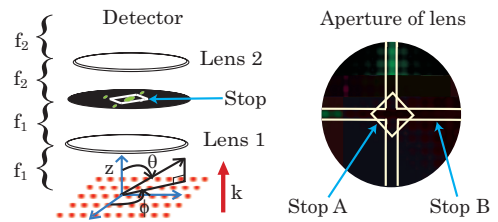}
\vspace{-0.3cm}
\caption{\label{fig:Graphic1} (Color online) Left: The arrangement for diffraction from a 2D optical lattice. The incident light propagates in the positive
$z$ direction and scatters off a regular array of atoms in the
$xy$ plane. The two lenses have focal lengths $f_1$ and $f_2$.
Right: View along the $z$-direction showing the intensity pattern of elastic scattering in the focal plane of lens 1 (with log scaling, as in Fig.~\ref{fig:Graphic2}), and the square (A), or cross-shaped stop (B), used to block the incident beam and most of elastically scattered light in central (zero order) diffraction peak (higher orders fall outside the range of angles collected by the lens). In the focal plane of lens 2 a photodiode, or similar detector, measures the intensity of the (unblocked) scattered light. }
\end{figure}

The temperature of fermionic atoms in optical-lattice experiments has been deduced
indirectly from the temperature of the trapped cloud of
atoms before turning up the lattice \cite{JOR08,SCH08}, and by detecting double-occupancy in a
two-species gas by converting atom pairs into molecules \cite{STO06,JOR08}. 
Other temperature measurements detected atomic shot-noise \cite{ROM06} 
or the sharpness of interference peaks \cite{CHI06} in absorption images
after a ballistic expansion. This existing technology
provided vital information about temperature but it has limitations, and
there is a clear need for new methods; e.g., detecting atomic shot noise proved inconclusive in some
superfluid/thermal lattice systems \cite{FOL05} and it was argued
that detecting superfluidity from the sharpness of interference
peaks can be ambigious as even a thermal gas may show misleadingly
sharp peaks \cite{DIE07}. Moreover, a range of phenomena
occur below the N\'{e}el temperatures \cite{neel} of these systems, e.g.,
antiferromagnetic ordering and superfluid pair hopping, but this
requires significantly more cooling than current
experiments \cite{JOR08}  further increasing the need for a method
to measure ultra-low temperatures. The diffraction of light from
regular arrays of atoms in an optical lattice resembles the powerful method of x-ray diffraction from
crystalline materials. The diffraction pattern reflects
the atomic lattice structure and the overall diffraction envelope the shape of atomic wavefunctions
at individual sites. Scattering into angles outside the diffraction
orders arises by
inelastic processes in which the phonon excitations of atoms in the
lattice absorb the recoil kicks from the scattered photons and
generate fluctuating shifts in the diffraction pattern which carry
information about thermal and quantum correlations of the atoms.
Light scattering from optical lattice systems was previously
studied theoretically for detecting particle number fractionalization
\cite{JAV03}, coupling of atoms to optical cavities \cite{MEK07} and
to polarization of light \cite{ECK08}.

We consider a fermionic atomic gas in an optical lattice illuminated by
light with the positive frequency component of the electric field
amplitude $\mathbf{E}_{\rm in}^+(\mathbf{r}) = \frac{1}{2}\xi \hat{\mathbf{e}}_{\rm in}
e^{\mathrm{i}\mathbf{k}\cdot\mathbf{r}}$, the polarization
$\hat{\mathbf{e}}_{\rm in}$, and the wave vector $\mathbf{k} =
k\mathbf{\hat{e}}_z$; see Fig.~\ref{fig:Graphic1}. In lattices the atom dynamics 
can be restricted to 1D (highly-elongated tubes) or 2D (pancake-shaped layers)
by optical confinement \cite{Smith}. For light passing through in an appropriate direction 
such a sample is optically thin and this makes light scattering a more suitable probe than in 3D samples.
In the limit of a large frequency detuning $\delta$ of the light
from the atomic resonance (compared to the Weisskopf-Wigner linewidth
$\gamma$), the dynamics of the electronically excited atomic
state may be adiabatically eliminated \cite{ruostekoski1}. The scattered light
amplitude $\mathbf{E}_{\rm sc}^{+}(\rv)$ then becomes proportional to the
transition amplitude of atoms between the initial and final
hyperfine (electronic ground) states, $g$ and $g'$, and we calculate
the scattered light at $\rv$ (in the far radiation field
\cite{comment}) by integrating over the atomic dipoles residing at
$\rv'$ \cite{ruostekoski1}:
\begin{equation}\label{Eq3}
\mathbf{E}_{\rm sc}^{+}(\rv) = C \mathbf{\Lambda}_{g'g}\int d^3r'
    e^{-\mathrm{i}\Delta\mathbf{k}\cdot\mathbf{r'}}\hat{\Psi}_{g'}^{\dag}(\mathbf{r'})\hat{\Psi}_{g}
    (\mathbf{r'})\,.
\end{equation}
Here
$\Delta \mathbf{k} = \mathbf{k'}-\mathbf{k} \simeq
k(\hat{\mathbf{n}}- \hat{\mathbf{e}}_{z})$ is the change of wave vector of light upon scattering
 ($|\Delta \mathbf{k}| = 2 k |\sin(\theta/2)|$) with
     $\hat{\mathbf{n}}$ being a unit vector in the direction of the detector from the sample,  $C = 3\xi e^{\mathrm{i}kr}\gamma/(4\delta k r)$,
$\mathbf{\Lambda}_{g'g} =
\hat{\mathbf{n}}\times(\hat{\mathbf{n}}\times
    \mathbf{d}_{g'e})( \hat{\mathbf{e}}_{\rm in}\cdot
    \mathbf{d}_{eg})/\mathfrak{D}^2$,
and $\mathbf{d}_{eg}$ =
$(\mathbf{d}_{ge})^{*} = \langle e|\mathbf{d}|g\rangle
=\mathfrak{D}\langle e|\sigma g\rangle\hat{\mathbf{e}}_{\sigma}^{*}
$ is the atomic transition dipole matrix element between $g$ and an
excited Zeeman state $e$, where
$\mathfrak{D}=(6\pi\hbar\eo\gamma/k^3)^{1/2}$ is the reduced dipole
matrix element and $\langle e|\sigma g\rangle $ are the Clebsch-Gordan
coefficients. Here we use the convention that repeated
indices ($g,e,\sigma$, etc.) are summed over.

Ultra-cold atoms only occupy
the lowest vibrational state of each lattice site and we can expand the
matter field in \eq{Eq3} in terms of the Wannier functions
$w_{g,j}(\rv)=w_g(\rv-\rv_j)$ localized at each site $j$
centered at $\rv_j$, i.e., we write $\hat{\Psi}_g(\mathbf{r}) =
\sum_{j}w_{g,j}(\rv)\hat{b}_{gj}$. Here $\hat{b}_{gj}$ denotes the
corresponding annihilation operator. We obtain for the
scattered light intensity $I(\mathbf{r}) = 2 \eo c\langle
\mathbf{E}_{\rm sc}^{-}\mathbf{E}_{\rm sc}^{+}\rangle$ which carries information
about the atomic correlations
\begin{equation}\label{Eq9}
I=
 {\sf M}_{g_2g_1}^{g_3g_4} \alpha(\unit) \textsl{B} \sum_{i,j}e^{\mathrm{i}\Delta
    \mathbf{k}\cdot(\rv_i-\rv_j)}\langle
    \hat{b}^{\dag}_{g_4i}\hat{b}_{g_3i}\hat{b}^{\dag}_{g_2j}\hat{b}_{g_1j}\rangle\,.
\end{equation}
Here $ {\sf M}_{g_2g_1}^{g_3g_4} = \Lambda_{g_3g_4}^* \Lambda_{g_2g_1}$,
and $B=I_{\rm in} (3\gamma/2\delta kr)^2$ with $I_{\rm in}=\eo c \xi^2/2$ being the incoming
light intensity. The Debye-Waller factor $\alpha(\unit)$ depends on the integrals of
$w_{g,j}$ and, for the case of a level-independent lattice potential,
it is obtained from the Fourier transform  of the lattice site density
$\alpha= |\int d^3 r\, e^{-\mathrm{i} \Delta \kv\cdot\rv} |w_g(\rv)|^2|^2$.

Equation~\eqref{Eq9} gives the scattered light intensity for an
arbitrary lattice system if the appropriate atomic correlation functions can
be evaluated. Here we shall apply it to a single-component, non-interacting FD gas in a 2D square lattice,  as illustrated in Fig.~\ref{fig:Graphic1}---inelastic scattering events
that carry information about thermal and quantum fluctuations
of atoms are mapped onto the light field, hence providing a sensitive {\it in situ} thermometer.

We assume that atoms initially occupy only one hyperfine level and that atoms scatter back to the same level (cycling transition).
Consequently, we drop any explicit reference to hyperfine levels $g$.
We consider a $\sigma^+$-polarized light beam, with wavelength 766.5\,nm, driving
$|1\rangle \equiv |4S_{1/2}; F = 9/2, m_F = 9/2\rangle
\rightarrow |2\rangle \equiv |4P_{3/2}; 11/2, 11/2\rangle$
transition of $^{40}$K, so that $\dv_{12}=\mathfrak{D}\hat{\bf e}_{+1}$ and
$ {\sf M}_{1,1}^{1,1} = (3+\cos 2\theta)/4$. In a single-component fermionic gas
$s$-wave scattering between ground state atoms is forbidden and, at low temperatures,
the atoms can be considered as non-interacting.

We consider a uniform 2D square lattice with periodicity $a$, potential 
$ V= sE_R\left[\sin^2(\pi x/a) + \sin^2(\pi y/a)\right]$, and the lattice-photon
recoil energy $E_R=\pi^2\hbar^2/(2ma^2)$. The atom
dynamics along the $z$-axis is assumed to be frozen out by strong confinement.
The Wannier functions $w_j(\rv)$ can be approximated by ground
state harmonic oscillator wave functions with frequencies
$\omega_z$ and $\omega_{x,y}=2\sqrt{s}E_R/\hbar$, obtained
by expanding the potential at the lattice site minimum. Then
$\alpha(\unit) = \prod_{i=x,y,z} \exp\{-((\Delta k_i)^2l_i^2/2 )\}$, with
$l_{i} = (\hbar/m\omega_i)^{1/2}$. The only contribution to
the Hamiltonian $\hat{H} = -J\sum(\hat{b}^{\dag}_i
\hat{b}_j +\hat{b}^{\dag}_j\hat{b}_i)$ arises from the hopping
$J\simeq 4 s^{3/4}e^{-2\sqrt{s}} E_R/\sqrt{\pi}$,
where the summation is over adjacent sites only.

The Hamiltonian is diagonalized in the
quasi-momentum space by $\hat{b}_j =(1/ N_s)\sum_{\mathbf{q}}u_{\mathbf{q}}\hat{a}_{\mathbf{q}}e^{\mathrm{i}\mathbf{q}\cdot
\mathbf{r}_j}$,
where the amplitudes $|u_\mathbf{q}|=1$, and $N_s$ is the number of
sites in each direction (both $x$ and $y$). The energy is given
$ {E_\mathbf{q}}_j = 4J\left\{ \sin^2\left(q_{xj}a/2\right) + \sin^2\left(q_{yj}a/2\right) \right\}$, with $\mathbf{q}_j = (q_{xj}, q_{yj}) = ( 2\pi/N_s a)(j_x, j_y)$,
where the integers may be chosen as $ j_{x,y} = -N_{s}/2,\ldots,(N_{s}/2-1) $.
We may now calculate the intensity in \eq{Eq9} by evaluating
$\langle \hat{b}^{\dag}_i\hat{b}_i\hat{b}^{\dag}_j\hat{b}_j\rangle$ using the phonon
operators, $\hat{a}_\qv$, to give
\beq \label{Eq18}
I/A= f^2 \mathfrak{A}_{\bar{\Delta \kv}} +  {1\over N_s^4}
\sum_{\qv,\qv'} \bar{n}_{\qv} (1- \bar{n}_{\qv'})\, \mathfrak{A}_{\bar{\Delta \kv}+\qv'-\qv}\,,
\eeq
where $A\equiv {\sf M}_{1,1}^{1,1} \alpha(\unit) \textsl{B}$,
$f = N/N_s^2 \leq 1$ is the filling fraction of the
lattice with $N$ being the total number of atoms, $\bar{n}_{\mathbf{q}}$
denotes the occupation numbers of the ideal FD distribution, and
\beq
\mathfrak{A}_{\bar{\Delta \kv}}=\prod_{j=x,y} {\sin^2\( N_s \bar{\Delta k}_j a/2\)\over 
\sin^2\( \bar{\Delta k}_j a/2\)}\,,
\eeq
is the diffraction pattern from a 2D square  array of $N_s\times N_s$ diffracting apertures; where $\bar{\Delta \kv}$ is the change in the wave vector
of light on the $xy$-plane.

\begin{figure}[h]
\includegraphics[width=0.45\columnwidth]{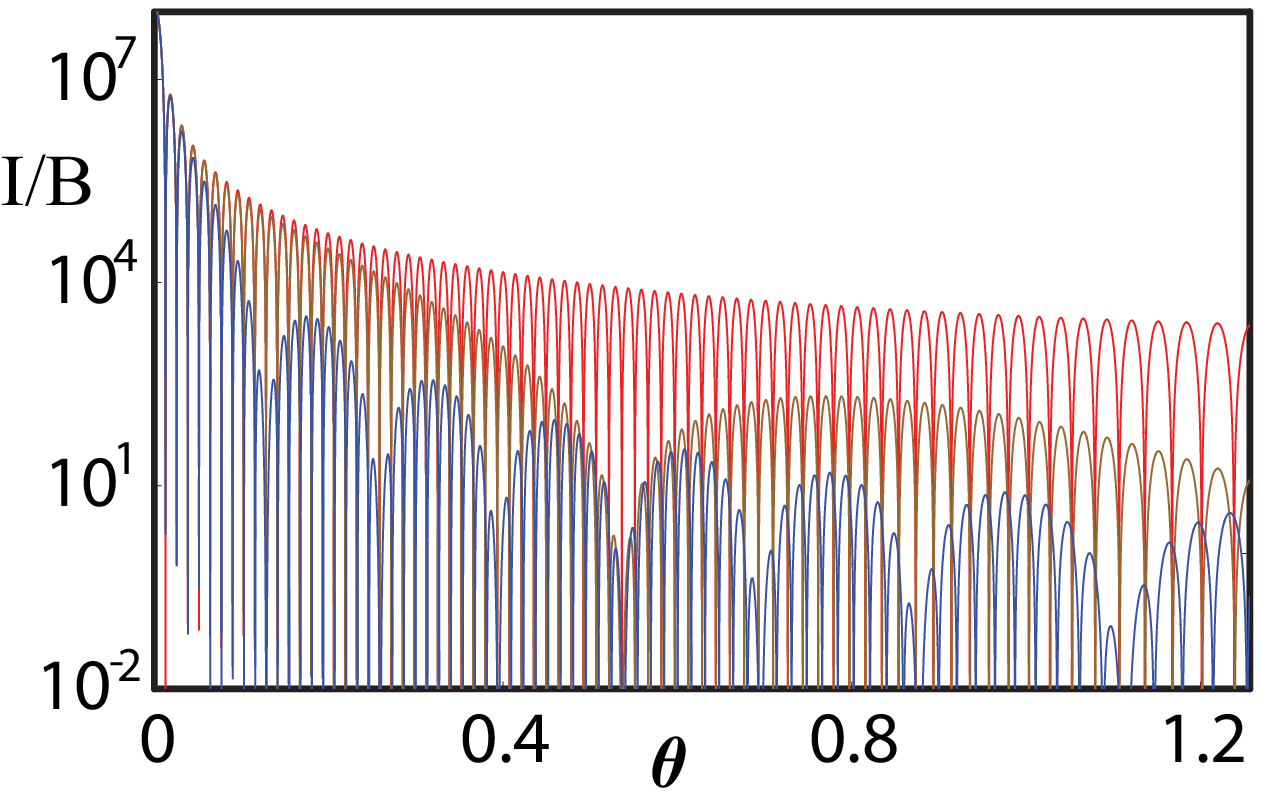}
\includegraphics[width=0.46\columnwidth]{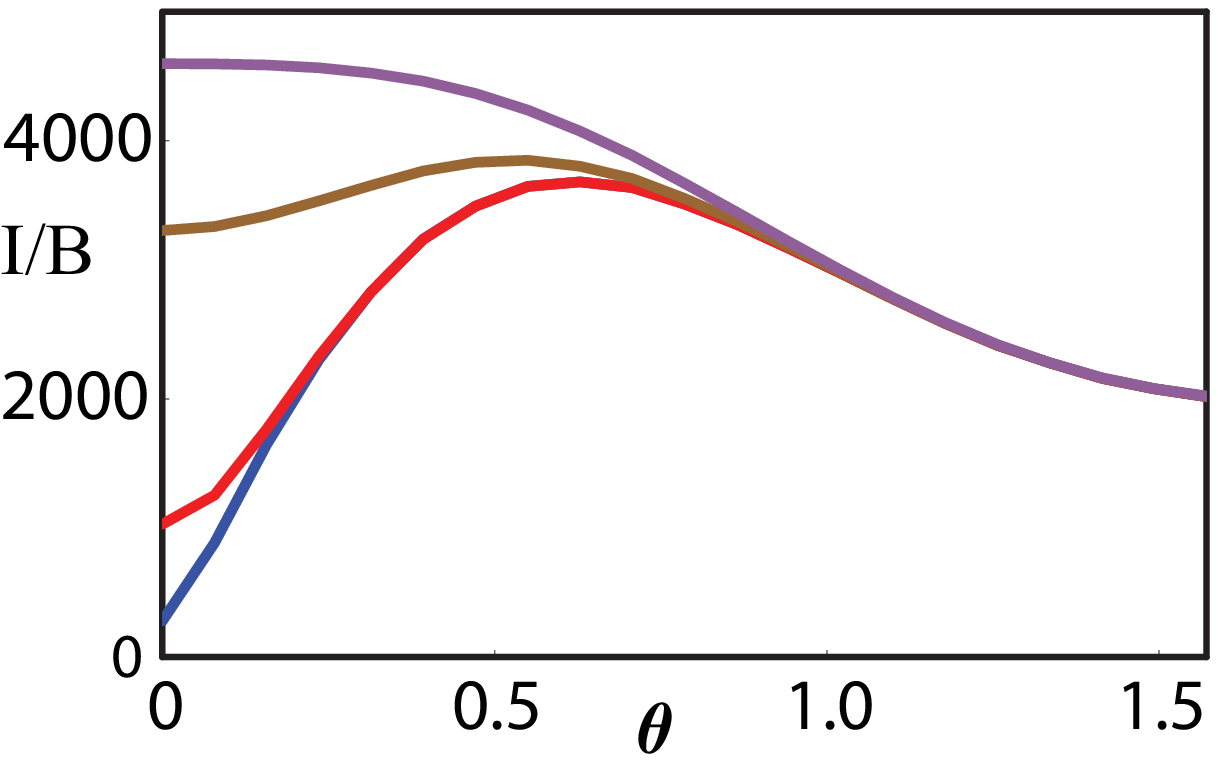}
\vspace{-0.4cm} \caption{\label{fig:Graphic2} (Color online) The angular
distribution of the intensity (in units of $B$ defined under
\eq{Eq9}) for elastic scattering (left on a logarithmic
scale) at scattering angles $\phi=0,0.025,0.1$\,rad (top to bottom); 
and for inelastic scattering (right on a linear scale)
showing the effect of fermionic statistics at temperatures $T/T_F =
0.01,0.05,0.25, 0.5$ (bottom to top) with $T_F = 64$\,nK and $\phi=0$.
Elastic scattering falls off faster in directions away from
the $x$ and $y$ axes (Fig.~\ref{fig:Graphic1}) therefore quantitative
comparison with the inelastic scattering requires integration over the
collected angles (Fig.~\ref{fig:Graphic4}).}
\end{figure}

The first term in \eq{Eq18}, proportional to $f^2$, is the elastic
scattering contribution where an atom scatters back to its original
c.m.\ state. It generates the diffraction pattern from a
non-fluctuating atom density; see Fig.~\ref{fig:Graphic1}. For a
large number of sites, this gives rise to sharp
diffraction peaks of small angular spread with significantly
weaker intensity between the orders. The second term in \eq{Eq18}
represents inelastic scattering where an atom is excited from a
quasi-momentum state $\qv$ and scatters to a {\it different} state
$\qv'$. It is this term that contains the effect of the FD
statistics, in both (a) the quasi-momentum distribution
$\bar{n}_{\qv}$ that obeys the quantum statistics, and (b) the
product of the occupation numbers that describes the Fermi
inhibition: scattering events in which an atom would recoil into an already
occupied state are forbidden by the Pauli exclusion principle.
Importantly, this inelastic term leads to scattering outside the
diffraction orders---phonon excitations lead to a deflection of the light since an atomic recoil $\Delta \qv=\qv'-\qv$
is associated with an equal and opposite change of photon momentum.
In a sequence of experimental realizations the values of
$\bar{n}_{\qv}$ fluctuate according to the FD statistics. Therefore
the second term in \eq{Eq18} generates, on the top of the diffraction pattern, an inelastically scattered light intensity that
also {\it fluctuates} during the sequence measurements.

To illustrate this method, we study light scattering from a
$N_s=150$ lattice of $^{40}$K atoms with $f=0.5$, $s = 7.8$,
$J=0.04\,E_R$, $a=0.4\,\mu$m, and $\phi = \pi/2$. The recoil 
component of the photon on the $xy$-plane is absorbed by
the atom quasi-momentum with $|\Delta \qv| = k \sin(\theta)$. Photon recoil to
higher bands on the $xy$-plane and in the $z$-direction is 
negligible (because we have taken $\hbar\omega_z\simeq 7.5E_R'$, where $E_R'$ is the probe
beam recoil energy, corresponding to  $l_z\simeq 63$\,nm and
$\omega_z/2\pi \simeq 63$\,kHz, and the energy band gap on the $xy$-plane 
is about 5$E'_R$ when the maximum energy absorbed by atoms due to a recoil kick
in the $xy$-plane is $E_R'$ \cite{comment2}). 
At $T=0$ the atoms fill the Fermi sea $\bar{n}_{\qv}=\Theta(q_F-|\qv|)$
and only inelastic scattering events for which the final state is
out of the Fermi sea are allowed. Thus all scattering events for
which $\pi/a>|\Delta \qv|>2q_F$ are allowed, but for small angles
$\sin(\theta) < 2q_F/k$ some recoil events would lead to an
already occupied state and are forbidden \cite{ruostekoski2}.
Thus inelastic scattering is strongly suppressed
by the FD statistics in the near-forward direction (or near the
diffraction maxima); here $2q_F/k\simeq1.36>1$ and it is at least partially
suppressed in all directions as seen at low $T$ in
Fig.~\ref{fig:Graphic2} (right).  At large angles
($\theta>0.5$\,rad) the diffraction pattern envelope is determined
by the internal atomic level structure and the lattice site wave
function (hence does not give information about the FD statistics).
Elastic scattering exhibits a diffraction peak in the forward 
direction which we block using a suitable stop (Fig.~\ref{fig:Graphic1}), 
together with any unscattered part of the incident beam; this light 
does not provide any information about the atom statistics.

\begin{figure}
\includegraphics[width=0.45\columnwidth]{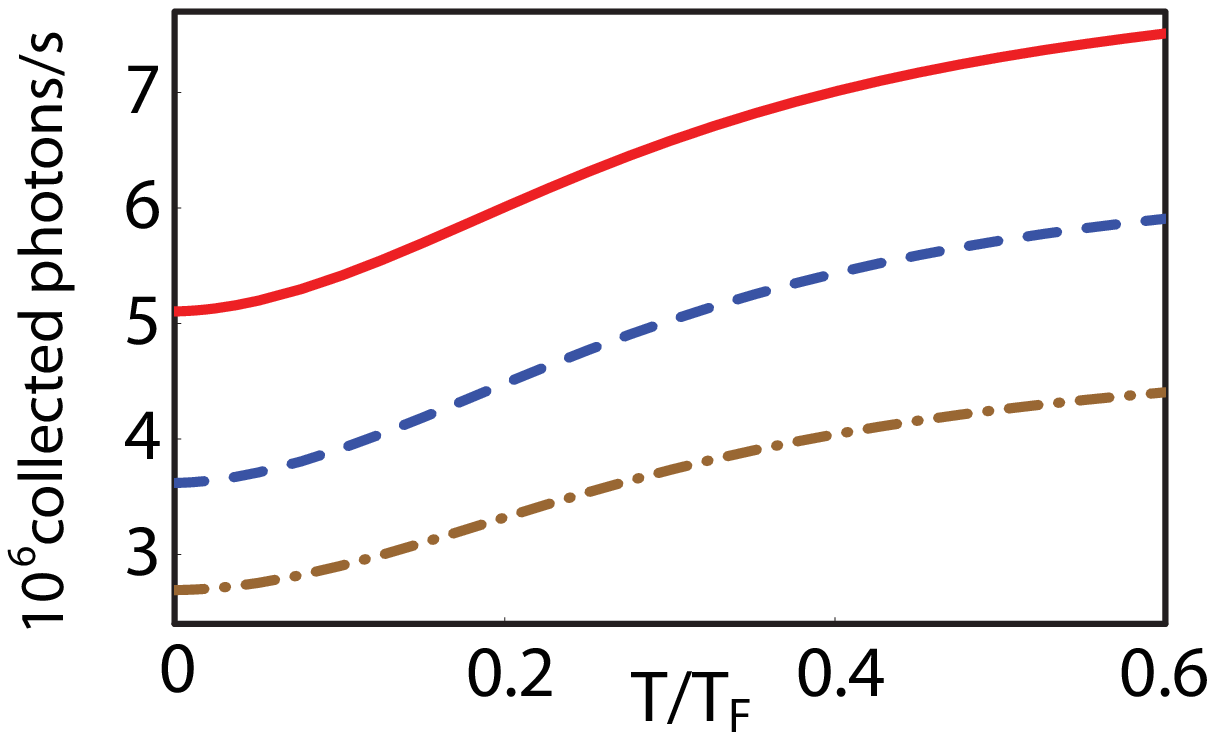}
\includegraphics[width=0.43\columnwidth]{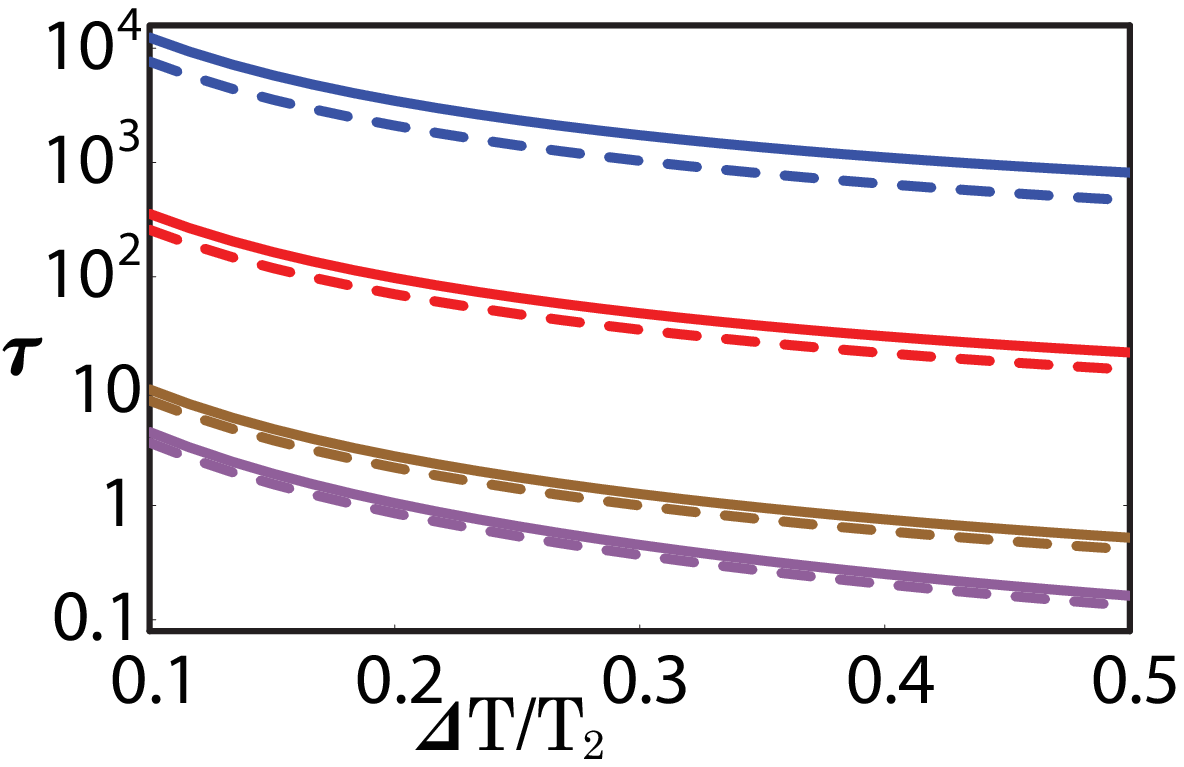}
\vspace{-0.4cm} \caption{\label{fig:Graphic4}(Color online) Left: The total number
of collected photons/second (elastic + inelastic) as a function of
temperature for a square stop (solid line), a cross-shaped stop 
of angular width 0.01\,rad (dashed line) and 0.05\,rad (dashed-dotted).
Right: The required number of experimental repetitions needed to measure temperature to certain relative precision ($\Delta T/T_2$), with
solid (dashed) lines for the square (0.01\,rad cross-shaped) stop and temperatures of $T/T_F= 0.02,0.05,0.15,0.25$ (top to bottom).}\end{figure}
In Fig.~\ref{fig:Graphic4} we show the number of collected
photons/second by the lens with the numerical
aperture NA=$\sin\theta_{\mathrm{max}}$=0.48 for
$I_{\rm in} = 5$\,Wm$^{-2}$ and $\delta = 20\gamma$.
We use a square stop of half-diagonal angular width of $0.05$\,rad with its diagonals
oriented along the $x$ and $y$ axes, and a cross-shaped stop, $|\theta_x|<\Theta , |\theta_y|<\Theta$, for $\Theta =0.01$ and
0.05\,rad to block most of the elastically scattered light without dramatically reducing the detection of inelastically scattered
photons \cite{elasticpart}.

In each inelastic scattering process an atom absorbs a recoil kick which changes its c.m.\ state.
This affects the temperature (heats the sample) and limits the maximum number of inelastic scattering events $W$ in a single experimental realization of
the lattice system to be  a small fraction $W/N$ of the total
number of atoms (otherwise the measurement perturbs the ground state significantly). Only a fraction $\eta(T)$ of all the inelastically
scattered photons are collected by the lens, so $N^{\mathrm{in}}_{\mathrm c}(T) = \eta(T)W$ is the number of
detected inelastically  scattered photons in each  realization (assuming 100\% detector efficiency).
In experiments the lattice system can be prepared and
measured $\tau$ times so that the total number of detected
photons is $\tau N_{\mathrm c}(T)= \tau [N^{\mathrm{in}}_{\mathrm c}(T) +N^{\mathrm{el}}_{\mathrm c}(T)]$, 
where $N^{\mathrm{el}}_{\mathrm c}$ denotes the number of detected, elastically scattered photons
in a single realization.
For far off-resonant light, the fluctuations of the number of scattered
photons are Poissonian, so to distinguish between two different
optical responses corresponding to temperatures $T_1$ and $T_2$ (with $T_2>T_1$),
the difference between the number of detected photons in the two cases
must be of order $\sqrt{\tau N_{\mathrm c}(T_2)}$.
Thus the number of repetitions of the measurement required for an uncertainty of $\Delta T = T_2 -T_1$ in the
determination of temperature is the value of $\tau$ which satisfies  $\tau(N_{\mathrm c}(T_2) - N_{\mathrm c}(T_1)) \simeq \sqrt{\tau
N_{\mathrm c}(T_2)}$.

For simplicity, we set $W/N = 0.1$. Figure~\ref{fig:Graphic4} 
shows the  number of experimental realizations of the lattice system
required to achieve an accuracy $\Delta T/T_2$,
e.g., $T/T_F = 0.1, 0.25$ can be measured within 5\% by
preparing the system 98 and 14 times respectively, providing a highly
sensitive thermometer for atoms \cite{Note}. In Ref.~\cite{JOR08} a Mott 
insulator state was observed at $T\simeq0.28T_F$. The fraction $\eta(T)$ 
of inelastically scattered photons collected by
the lens is large in Fig.~\ref{fig:Graphic4} (varying from 18 to 37\%). This arises because 
(a) the intensity in
the forward direction is twice the intensity in the transverse
direction for $\sigma$-polarized light (a well-known feature of the
Zeeman effect); and (b) the size of the lattice site wave function
generates an overall envelope for the diffraction pattern that
suppresses radiation at large angles. We found that the signal-to-noise ratio is not improved with a larger aperture lens 
(NA=0.75), but would be increased by using a stack of several identical 2D lattice layers, provided that the system remains optically thin.
\begin{figure}[t]
\includegraphics[width=0.45\columnwidth]{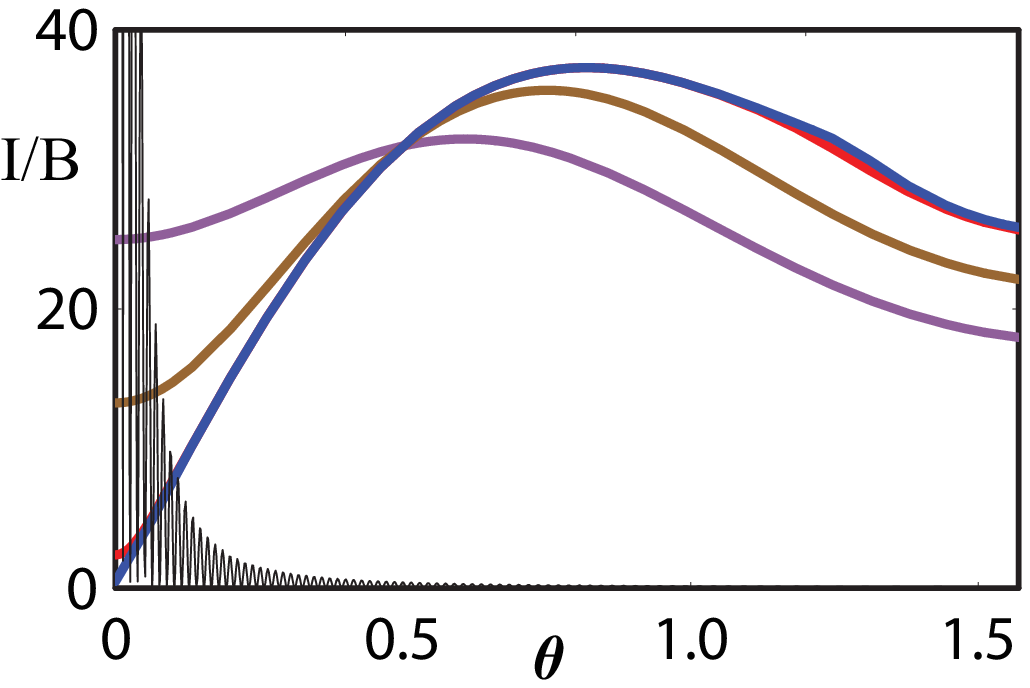}
\includegraphics[width=0.45\columnwidth]{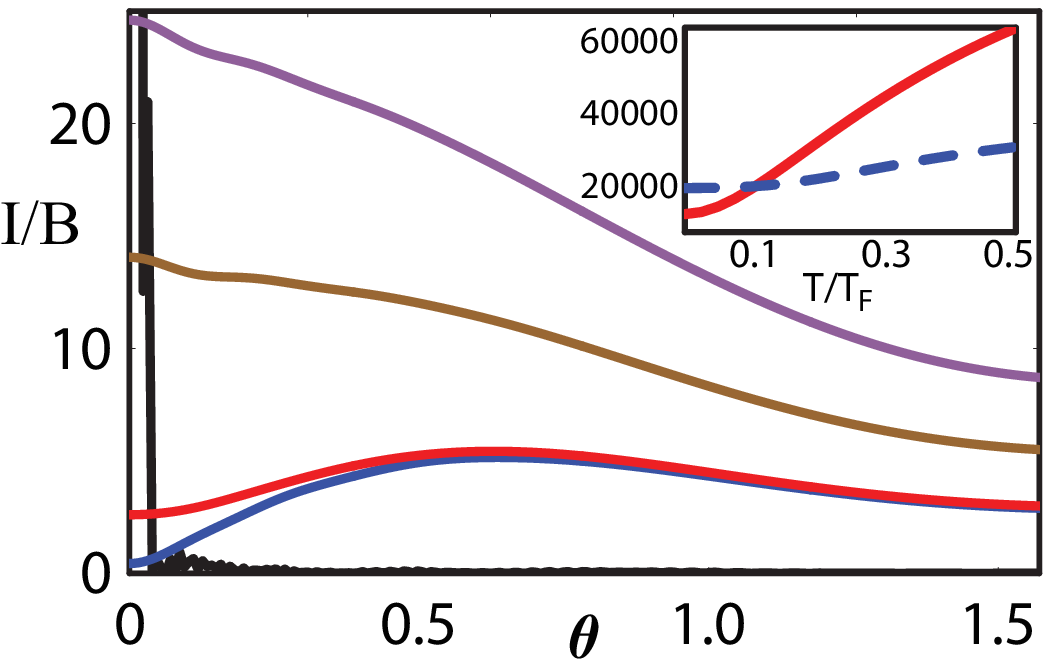}
\vspace{-0.4cm} \caption{\label{fig:Graphic5}
(Color online) The angular distribution of the scattered
light intensity (at $\phi=0$) from a 1D lattice (left), and the same lattice with an
additional harmonic potential in which $l_h = 2\,\mu$m (right). The rapidly oscillatory
curves on both sides represent elastic scattering (on right
for $T = 0.5\,T_F$) and the smooth curves correspond to inelastic
light at $T/T_F = 0.01, 0.05, 0.25, 0.5$ (bottom to top
at $\theta= 0$). The inset shows the temperature dependence of
the number of collected photons/second (elastic 
+ inelastic) for a stop with $|\theta_x|\leq \Theta$ for the two cases: no trap, NA=0.48,
$\Theta = 0.047$\,rad (dashed line); and atoms in a potential with $l_h = 2\,\mu$m, NA$=0.8$, $\Theta = 0.086$\,rad (solid line).}\end{figure}

In typical experiments atoms are confined in an optical lattice plus harmonic trap---to check that
this does not render our method inaccurate we carried out calculations for a 1D lattice system with a harmonic
potential of frequency $\Omega$.  This affects the plane
wave basis used in \eq{Eq18} when the variation of the trap energy over the sample is greater than the hopping
energy: $m\Omega^2(N_s a)^2 \agt J$; which for our parameters occurs when 
$\Omega \agt 2\pi \times 7$\,Hz ($l_h = \sqrt{\hbar/m\Omega}\alt 6\,\mu$m). 
In a trapping potential we diagonalize the Hamiltonian $H = \sum_j [\zeta j^2 \hat{b}_j^\dagger
\hat{b}_j-J(\hat{b}_j^\dagger \hat{b}_{j+1}+\hat{b}_{j+1}^\dagger\hat{b}_j)]$,
with $\zeta = (a^2/\pi l_h^2)^2E_R$, and use the eigenfunctions
as a new basis to evaluate \eq{Eq9}. For the same
parameters as in the 2D case ($N_s = 150$ with the lattice
along the $x$ axis and $l_{y,z} = 63$\,nm) we find that the angular
distribution of the scattered light for a trap with $l_h = 6\,\mu$m
is modified very slightly (less than 2\% difference at any point), 
whereas the results for an additional potential in which $l_h = 2\,\mu$m are shown in Fig.~\ref{fig:Graphic5}. 
The temperature
variation in the trap is notably stronger than for a translationally invariant lattice because the inelastic scattering depends on temperature
at all angles. Thus the temperature sensitivity of the measured signal is {\it enhanced}.

In conclusion, we propose an efficient optical thermometer
for fermionic atoms in a lattice.  The general scattering formula \eqref{Eq9} may easily be applied also to
multi-species systems \cite{CHI06,JOR08,SCH08}, provided that the atomic correlation
functions can be evaluated.

ABD (via QIP IRC) and JR acknowledge financial support from the EPSRC.


\begin{thebibliography}{24}
\bibitem{comment2} Such recoil events would also correspond to photons
scattering at $\theta=\pi/2$ and which would not be collected by the lens.
\bibitem{elasticpart} For the cross-shaped stops of 0.01 and 0.05\,rad the collected 
elastic part is less than 7\% and 0.5\%, respectively,
at $T/T_F=0.5$. The smaller stop
gives a better signal. The stops also efficiently block any unscattered
remnant of the incident beam.
\bibitem{Note} Elastic scattering depends only on the total atom number
in the lattice system (constant during each experimental realization) 
and does not alter the state of the atoms, as long
as the intensity variation of the incoming light beam is
small over the atomic sample--this is analogous to the
effect of a quantum non-demolition measurement.
\end{thebibliography}
\end{document}